\documentclass[11pt]{article}
\usepackage{pgfplots}

\usepackage{graphicx}
\usepackage{amsmath}
\usepackage{pgfplots}
\usepackage{caption}
\usepackage{subcaption}
\usepackage{float}
\usepackage{float}

\begin{document}

\def\vghl{\color{red}}
\title{\ Accounting for anomalous velocity curve in galaxies as a possible frame dragging effect}
\author{Investigator: Amey Gupta\thanks{E--mail : ameyg.2021@gsuite.aditi.edu.in}\\ 
Mentor: Daksh Lohiya\thanks{E--mail : dlohiya@physics.du.ac.in}\\
  Mallya Aditi International School, Bengaluru, Karnatakai\\
  \& \\
  D.A.M.T.P., University of Cambridge, Cambridge CB3 0DT, UK}

\renewcommand {\today}{}

\maketitle
\begin{abstract}
We investigate the possibility of accounting for the observed anomalous velocities of stars in
galaxies to be a result of a dynamic prescription for having a galaxy
generate and prescribe its own inertial frame. It is demonstrated that within observational uncertainties,
an overall rotation of an inertial frame dragged by a galaxy is a comfortable fit to the observed
anomalous velocity curves. A case for looking for a natural law for prescribing an inertial frame by a mass
distribution over a finite (galaxy size) range is made out.

\end{abstract}

\section{Anomalous Velocities of Stars in Galaxies}

 It is now well accepted that the variation of rotational speed of stars with the distance from the center of
the host galaxy cannot be accounted in terms of the distribution of baryonic matter alone. For disc galaxies, the
rotational velocity approaches a constant, instead of displaying the expected Newtonian  fall off as $V \approx r^{-1/2}$ \cite{Rubin,Bosma}. Similar discrepancy is observed in galaxies over a wide range of morphologies. To account for this discrepancy in terms of a possible ``dark matter'' component has now become the state of the art. Of late,
a strong correlation between the amount of such dark matter, required to be present in a galaxy, and the baryonic
component, has been reported \cite{Stacy}. This suggests either (a) a search for a new dark matter sector physics that could account for the observed correlation with baryons; or (b) The correlation could be a 
result of new dynamical laws.\\

In this letter we report the possibility of an inertial frame drag that could account for the observed features of
rotation curves.\\

Classical Newtonian mechanics does not have any dynamic law that prescribes an inertial frame from a
given distribution of particles.
Newton was well aware of this issue and skirted it by simply identifying an inertial frame with {\it{''the frame of
the fixed stars''}}. Ernst Mach expressed the inadequacy of such a prescription of an inertial frame. Indeed, classical 
mechanics has no prescription to deal with a hypothetical system consisting of a finite number of interacting particles, distributed over a finite domain in an otherwise empty universe. At any given instant, with the particle positions and velocities prescribed, classical mechanics has no prescription that allows one to specify an inertial frame. Such a frame would be one in which an accelerating particle would experience a fictitious force. \\

One can conceive of many ''matter of principle'' problems in configurations that can not be addressed to for such a localized  distribution of a finite number of particles. Take for example a system consisting of two spheres that are rotating with respect to each other, along an axis joining their centres, in an otherwise empty universe. The two spheres are thus mutually rotating. Classical physics alone does not offer a reply to a query: {\it{On which of the two spherical
surfaces would an observer feel a centrifugal force?}} Newton would want to refer to some distant star - which is absent in the stated problem. According to Mach's Principle, in effect one must specify some dynamic prescription that would allow one to determine the inertial frame for a given distribution of matter. We entertain such a possibility and a further that the interaction necessary to dynamically prescribe an inertial frame could have a finite range of the order of the size of a typical galaxy. We define this range as "{\it{the inertial range}}". Such a galactic system would thus have an embedded inertial frame that could have a significant drag component along with the galaxy itself.\\

In effect, to the expected Newtonian velocity curve, one would have to add the overall rotation of the embedded inertial frame. To a first approximation, if the size of a galaxy is much less than  {\it{the inertial range}}, the 
net effect would be to add a component proportional to the fixed angular speed of the embedded inertial frame. \\

We considered examples of galaxies of different morphologies: (i) Bulge - Dominated Spirals; (ii) Disk - dominated 
Spirals; and (iii) Gas - Dominated Dwarfs as described in \cite{Stacy}. The observed and expected velocity curves are 
reproduced in Figure (1) for the Bulge Dominated and the Disk Dominated Spirals.
\\

We discover that to a very good approximation, the difference between the observed and expected velocities vary linearly with distance - right from the core to a significant distance beyond it. Small deviations at large distances could by symptomatic of the possibility of the distanced from the centers becoming comparable to the 
{\it{ inertial range}}. Figure (2) describes the empirical linear variation of the anomalous velocities with distance.\\

\begin{figure}[H]
\begin{center}
\begin{tabular}{c c}
\centering
\includegraphics[width=70mm, height=58mm]{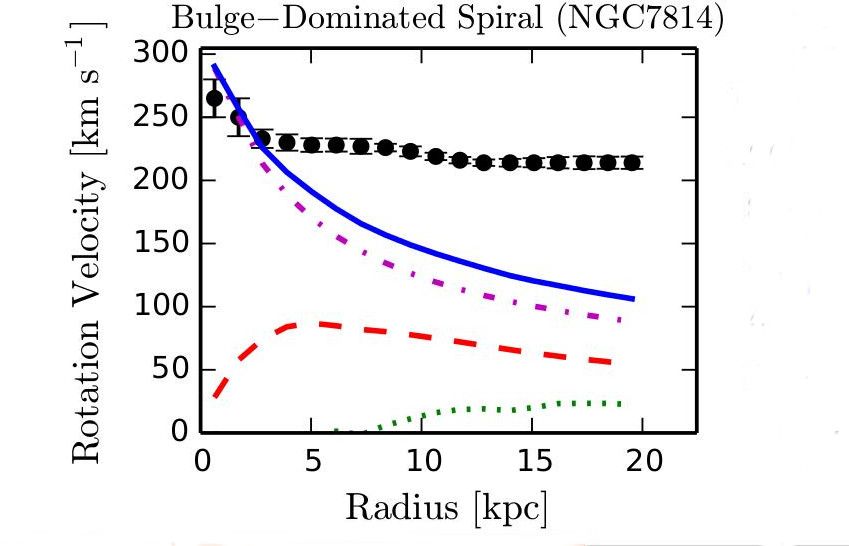}
\includegraphics[width=52mm, height=58mm]{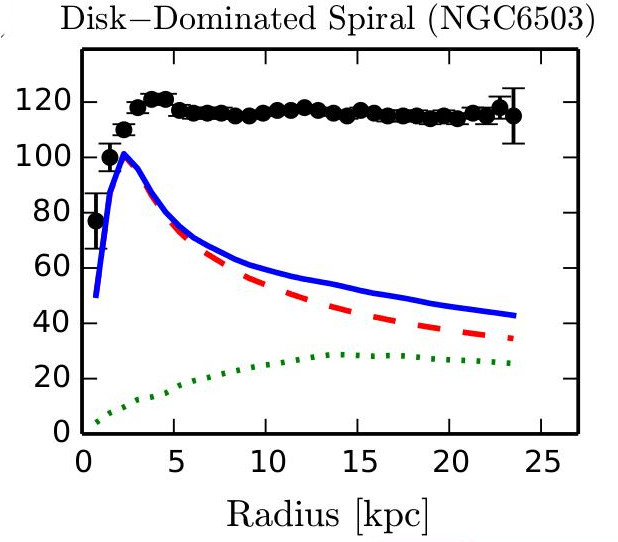}
\end{tabular}
\label{fig1}
\caption{ Left panel shows the variation of velocity of a star in a Bulge Dominated Spiral Galaxy with distance. The right panel displays the same for a Disk - Dominated Spiral Galaxy: taken from Stacy et al \cite{Stacy}.}\label{fig:b0}
\end{center}
\end{figure}

\begin{figure}[H]
\begin{center}
\begin{tabular}{c c}
\centering
\includegraphics[width=60mm, height=58mm]{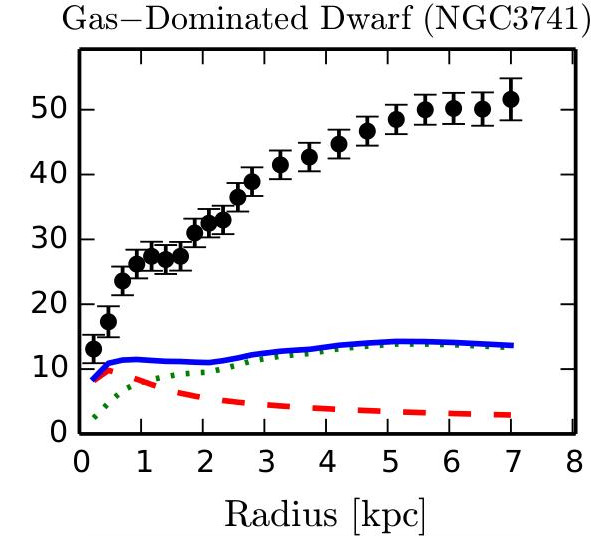}
\includegraphics[width=60mm, height=60mm]{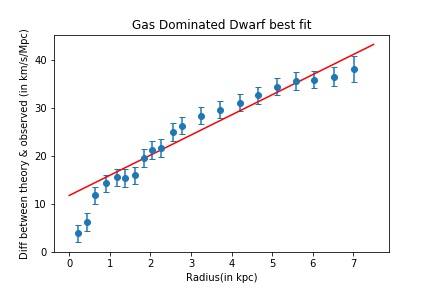}
\end{tabular}
\label{fig2}
\caption{ Left panel shows the variation of velocity of stars in a Gas Dominated Galaxy\cite{Stacy}. The right panel displays the best linear fit variation with distance for the discrepancy of the observed and the theoretical curves. This would represent the rotation of an embedded inertial frame.}
\end{center}
\end{figure}

\begin{figure}[H]
\begin{center}
\begin{tabular}{c c}
\includegraphics[width=60mm, height=58mm]{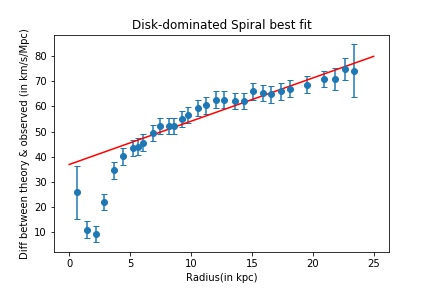}
\includegraphics[width=60mm, height=58mm]{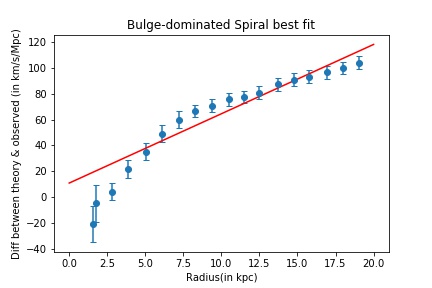}
\end{tabular}
\label{fig3}
\caption{ Left panel displays the best linear fit for variation  with distance for the discrepancy between the observed and theoretical curves for the Disk Spiral Galaxies while the right panel displays the same for the Bulge Spiral Galaxies. This would represent the rotation of an embedded inertial frame.}
\end{center}
\end{figure}

\clearpage

\section{Discussion and Conclusion}

The empirical observations reported in this letter make out a case for a search of a formalism that could specify an inertial frame from the given distribution of matter. The linear variation of the anomalous velocity discrepancy 
of stars in galaxies with distance suggests that the anomaly could be a result of a system as large as a galaxy having
a dedicated embedded inertial frame. Further, the fall off of the velocity discrepancy at large distances suggests that the {\it{the inertial range}} could be comparable to the size of a typical galaxy.\\

\vskip 0.5 cm

{\bf Acknowledgments}

One of us (A.G.) is grateful to the Department of Physics and Astrophysics for allowing him to research with his mentor Prof. Daksh Lohiya under the INSPIRE and INNOVATE programme. 

\vskip 1cm



\bibliography{plain}

\end{document}